# Breakdown-Limited Endurance in HZO FeFETs: Mechanism and Improvement Under Bipolar Stress

**Kasidit Toprasertpong[1*], Mitsuru Takenaka[1], Shinichi Takagi[1]**

[1] Department of Electrical Engineering and Information Systems, the University of Tokyo, Tokyo, Japan

**\* Correspondence:**
Kasidit Toprasertpong
toprasertpong@mosfet.t.u-tokyo.ac.jp

**Keywords: Ferroelectrics, MOSFET, reliability, oxide breakdown, substrate hole current.**

**Abstract**

Breakdown is one of main failure mechanisms that limit write endurance of ferroelectric devices using hafnium oxide-based ferroelectric materials. In this study, we investigate the gate current and breakdown characteristics of $Hf_{0.5}Zr_{0.5}O_2$/Si ferroelectric field-effect transistors (FeFETs) by using carrier separation measurements to analyze electron and hole leakage currents during time-dependent dielectric breakdown (TDDB) tests. Rapidly increasing substrate hole currents and stress-induced leakage current (SILC)-like electron currents can be observed before the breakdown of the ferroelectric gate insulator of FeFETs. This apparent degradation under voltage stress is recovered and the time-to-breakdown is significantly improved by interrupting the TDDB test with gate voltage pulses with the opposite polarity, suggesting that defect redistribution, rather than defect generation, is responsible for the trigger of hard breakdown.

## 1 Introduction

$HfO_2$-based ferroelectric thin films have been actively employed in recent electron device research thanks to their CMOS compatibility, established know-how on the fabrication process, and high scalability of thickness to 10 nm or lower (Böscke et al., 2011a; Müller et al., 2012; Park et al., 2015; Migita et al., 2018b; Kim et al., 2018; Tan et al., 2021; Toprasertpong et al., 2022a; Schroeder et al., 2022). Ferroelectric field-effect transistors (FeFETs) with $HfO_2$-based ferroelectric thin films as gate insulators have received considerable attention, not only because of the maturity of the $HfO_2$ deposition technology in the advanced transistor process, but also because of their low energy consumption, high speed, and satisfactory retention during their operation. $HfO_2$-based FeFETs have been investigated as promising devices for low-power nonvolatile memory (Böscke et al., 2011b; Trentzsch et al., 2016; Dünkel et al., 2017; Florent et al., 2018a; Müller et al., 2021) and non-von Neumann computing applications (Jerry et al., 2018; Dutta et al., 2022; Matsui et al., 2021; Luo et al., 2022; Toprasertpong et al., 2022b).

Despite their excellent properties, one of the most crucial issues to be dealt with towards the practical use of $HfO_2$-based FeFETs is the write endurance. There are two major mechanisms that have been reported to determine the write endurance of FeFETs: the memory window narrowing and gate dielectric breakdown. The memory window narrowing refers to a phenomenon where a separation of the threshold voltages of the two states (high and low threshold voltage states) becomes gradually smaller and eventually becomes zero after certain operating cycles. The polarization states are no longer able to be read out through threshold voltages and FeFETs lose a capability as memory devices. On the other hand, gate dielectric breakdown refers to a situation where the gate insulator experiences hard breakdown under a certain amount of electrical stress. Hard breakdown makes gate insulators conductive, electrically connects the gate and channel, and causes FeFETs to lose their function as field-effect transistors.

Memory window narrowing and gate dielectric breakdown originate from different physics and occur almost independently; therefore, the write endurance of FeFETs, i.e. a number of write operations before failure, is determined by the mechanism that leads to earlier failure. The dominant mechanism depends on the device property and the operation scheme of each specific device and application. Write endurance of state-of-the-art FeFETs is typically dominated by the memory window narrowing (Böscke et al., 2011b; Yurchuk et al., 2016; Trentzsch et al., 2016; Dünkel et al., 2017; Florent et al., 2018a; Gong et al., 2018) because of the presence of large density of trapped charges in the vicinity of the interfacial layer (IL) between $HfO_2$ and Si (Toprasertpong et al., 2019; Toprasertpong et al., 2020a), while there are only a few reports showing that endurance of FeFETs is limited by gate dielectric breakdown (Ni et al., 2018; Peng et al., 2021). That is, the FeFET operation so far usually reaches failure because of memory window narrowing before gate dielectric breakdown occurs; thus, there is still a poor understanding of the gate dielectric breakdown mechanism in $HfO_2$-based FeFETs. On the other hand, a lot of effort has been put on the material and device-structure engineering such that there have already been some reports in recent years demonstrating FeFET memory devices with remarkably suppressed memory window narrowing (Sharma et al., 2020; Yan et al., 2020; Tan et al., 2021; Liao et al., 2022). In such devices with suppressed memory window narrowing, gate dielectric breakdown may become a dominant mechanism that limits write endurance and play a crucial role in device reliability. Furthermore, there are some applications of FeFETs using new-concept computing that are insensitive to memory window narrowing, such as reservoir computing (Nako et al., 2022). In such applications, gate dielectric breakdown will be a dominant endurance-limiting mechanism. Therefore, gaining an understanding of the mechanism of gate dielectric breakdown is important to improve the overall write endurance characteristics of $HfO_2$-based FeFETs.

In this study, we investigate the breakdown characteristics and the stress-induced degradation behavior as well as the underlying physical mechanism in $Hf_{0.5}Zr_{0.5}O_2$ (HZO)/IL/Si FeFETs. The carrier separation measurement and interrupted stress for time-dependent dielectric breakdown (TDDB) evaluation are employed to analyze the physical mechanism underlying gate dielectric breakdown.

## 2    Sample Preparation

The process flow is shown in Figure 1A. We fabricated *n*-channel non-ferroelectric FETs (called here as nonferro-FET) with a paraelectric $HfO_2$ gate insulator and FeFETs with a ferroelectric HZO gate insulator on p-type Si substrates with a moderate doping concentration of $4\times10^{15}$ cm$^{-3}$. After the source and drain (S/D) regions were doped by phosphorus ion implantation and annealed to activate dopants, the Si substrates were cleaned by hydrochloric-peroxide mixture (HPM)-last cleaning process to grow a high-quality $SiO_2$ IL (Toprasertpong et al., 2020). For FeFETs, 10-nm-thick ferroelectric HZO was deposited by atomic layer deposition (ALD) using at using tetrakis(ethylmethylamino)hafnium (TEMAH), tetrakis(ethylmethylamino)zirconium (TEMAZ), and $H_2O$ at 300°C. For nonferro-FETs, 10-nm-thick $HfO_2$ was deposited in a similar way but without TEMAZ. TiN was deposited as gate metal by sputtering and Al:Si was deposited as S/D contacts by thermal evaporation. Samples were annealed at 400°C for 30 s in a $N_2$ atmosphere to crystalized the ferroelectric phase in FeFETs. The nonferro-FETs were also annealed at the same condition. Except the ALD step, both samples were processed simultaneously in the same chamber to ensure the same device condition. Figures. 1B and 1C show transmission electron microscopic (TEM) images of the gate stacks of a nonferro-FET and a FeFET, respectively, indicating that HZO was crystallized whereas $HfO_2$ remained amorphous. The IL thickness was similar in the both samples.

## 3  Results and Discussion

### 3.1  Band diagram and breakdown position

Before we discuss the experimental results of the leakage and breakdown behaviors, we examine the band diagram of the HZO (10 nm)/IL (0.7 nm)/Si gate stack and the possible gate leakage path. Figure 2A depicts an example of an ideal band diagram of the HZO/IL/Si gate stack at 3 V when HZO has ferroelectric polarization of 10 μC/cm². Due to high ferroelectric polarization, most literature considers a band diagram with a strong electric field across the IL, which significantly pulls down the band position HZO, as shown in Figure 2A (Müller et al., 2016; Yurchuk et al., 2016; Gong et al., 2018; Peng et al., 2021; Mulaosmanovic et al., 2021). In such a case, the breakdown of the IL is supposed to determine the gate dielectric breakdown of FeFETs. However, it has been reported that a large density of trapped charges near the HZO/IL interface electrically screens the polarization and suppresses the electric field across the IL (Toprasertpong et al., 2019; Toprasertpong et al., 2022c). Figure 2B depicts the band diagram with ferroelectric polarization of 10 μC/cm² and 90% (Ichihara et al., 2020) of induced electrons are trapped at the HZO/IL interface. It can be seen that the band of HZO is not at such a low energy position. This fact indicates that electrons have to tunnel through a thick HZO layer and thus the breakdown of HZO is necessary to describe the gate breakdown failure of FeFETs.

### 3.2  Device characteristics

The $I$-$V_g$ characteristics of the nonferro-FET and FeFET are shown in Figures 3A and 3B, respectively, for gate current $I_g$, drain current $I_d$, source current $I_s$, and substrate current $I_{sub}$. A gate length $L$ is 10 μm and a gate width $W$ is 100 μm. As expected, the nonferro-FET exhibits the $I_d$-$V_g$ characteristics with clockwise hysteresis, which is a feature of electron trapping during $V_g$ scans. On the other hand, the FeFET exhibits counterclockwise hysteresis, which is a feature of ferroelectricity, with a memory window of approximately 1.8 V. Comparison of the $I$-$V_g$ characteristics of the nonferro-FET and FeFET indicates interesting features on $I_g$ and $I_{sub}$. Gate current $I_g$ in the HZO FeFET is much larger by several orders of magnitude than in nonferro-FETs having $HfO_2$ with a similar physical thickness. This can be understood from the fact that the poly-crystallinity and a lot of defects such as oxygen vacancies in HZO can promote the gate leakage current, as shown in Figure 3C. It is also found that the substrate current $I_{sub}$ in the FeFET rapidly increases by four orders of magnitude in a narrow range of $V_g$ = 3.6 V to 4.0 V during the forward $V_g$ scan, which is in the same range that $I_g$ also increases rapidly by two orders of magnitude. This finding suggests that a study of the behavior of $I_{sub}$ would be helpful in understanding the behavior of the gate leakage and gate dielectric degradation. The nonferro-FET in Figure 3A does not exhibit this $I_{sub}$ behavior.

### 3.3  Carrier Separation Measurements

# Breakdown-Limited Endurance in HZO FeFETs: Mechanism and Improvement Under Bipolar Stress

Carrier separation measurements (Eitan et al., 1983; Weinberg et al., 1985) were carried out to analyze the behavior of gate leakage and gate dielectric degradation. The electrical measurement tool (Keysight B1500A with high-resolution source/monitor unit modules) was connected with FETs in a way shown in Figure 4A, where $V_d = V_s = V_{sub} = 0$. The current detected at the S/D terminal corresponds to the electron component of gate current, denoted by $I_e$, while the current detected at the substrate corresponds to the hole component, denoted by $I_h$. When $V_g$ is larger than the threshold voltage, $I_e$ corresponds to the tunneling current of inversion electrons from the Si substrate to the gate, whereas $I_h$ corresponds to the sum of the tunneling current of valance-band electrons in the Si substrate to the gate (Weinberg et al., 1985; Schuegraf et al., 1994b; Shanware et al., 1999) and the tunneling back current of holes from the gate to the Si substrate (Schuegraf et al., 1994a; Schuegraf et al., 1994b; Kobayashi et al., 1995), as illustrated in Figure 4B.

The results of the carrier separation measurements are shown in Figures 4C and 4D for the $HfO_2$ nonferro-FET and HZO FeFET, respectively. In these measurements, $V_g$ of pristine samples was scanned from 0 V to the positive voltage where breakdown occurs. It can be seen that tunneling of inversion electrons is the main contribution of $I_g$ for both the nonferro-FETs and FeFET. $I_h$ is found to be under detection limit in a low $V_g$ regime, but it rapidly increases at $V_g$ close to the breakdown voltage. The breakdown voltage $V_{BD}$ of the nonferro-FET is approximately 5.2 V, whereas the FeFET reaches hard breakdown much earlier at approximately $V_{BD} = 4.1$ V. Earlier breakdown is contributed to more defects in HZO than those in $HfO_2$, in agreement with larger gate current shown in Figures 3A and 3B. Hard breakdown of the nonferro-FET occurs at comparatively low $I_h$, whereas $I_h$ of HZO FeFET keeps noisy until very high level of $I_h$. After breakdown, the electrical properties of the gate insulators of both the devices become ohmic and dominated by electron current, as shown in Figures 4E and 4F.

The band alignments are shown in Figures 4G-4I. At small $V_g$, it is clear from the band alignment that electrons in the conduction band of Si can easily tunnel to the gate. At $V_g$ in the mid-range, both electrons in the valence band and holes generated at the gate can tunnel more easily, resulting in increasing $I_h$. At large $V_g$, an electric field across HZO is so large that hole tunneling back can reach the valence band of HZO, resulting in large $I_h$. Increasing hole tunneling back consequently causes breakdown in the gate insulator, as the hole tunneling back is known to be the main cause of damage in the gate insulator (Schuegraf et al., 1994a; Schuegraf et al., 1994b; Takayanagi et al., 2001).

Results of repeated measurements of $I_e$ and $I_h$ in a $V_g$ scan range of -2 V to 4 V are shown in Figure 5. It is interesting that rapidly increasing $I_h$ and $I_e$ at $V_g > 3.5$ V in the FeFET, together with noisy signals before breakdown, are recovered during the $V_g$ backward scan, resulting in repeatable $I_h$-$V_g$ and $I_e$-$V_g$ characteristics. These results imply that, although rapidly increasing $I_h$ is an indication that breakdown is going to be triggered, the permanent degradation still does not occur yet in this condition and occurs when $I_h$ increases in a step-wise manner, which can be observed in Figure 4D at $V_g = 4.1$ V.

The analysis above suggests that $I_h$ is a convenient indicator for determining appropriate operating range of $V_g$. Figure 6 shows the $I$-$V_g$ characteristics of the FeFET when $V_g$ was kept below 3.5 V. In this $V_g$ range, the ferroelectric hysteresis can still be achieved with a satisfactory memory window of 1.7 V while $I_h$ is suppressed to under the detection limit. Note that $I_{sub}$ at negative $V_g$ is due to gate-induced drain leakage (GIDL), which is unrelated to gate leakage currents. Although $I_h$ does not necessarily imply to device degradation as discussed in Figure 5, hole tunneling back is flowing and leads to a higher probability that breakdown is triggered; therefore, the operating condition with high $I_h$ should be avoided. The reliability of FeFETs operating in this way is notably improved and we cannot observe breakdown under electrical stress for a practically long time ($> 10^5$ s).

### 3.4 Time-dependent dielectric breakdown: Constant voltage stress and interrupted test

TDDB tests with a carrier separation setup were carried out to gain more insights into the breakdown behavior of FeFETs. $I_e(t)$ and $I_h(t)$ under constant voltage stress (CVS) as a function of stress time $t$ are shown in Figures 7A and 7B for nonferro-FETs and FeFETs, respectively. Both $I_e(t)$ and $I_h(t)$ of the FeFET increase with time, which is in the opposite direction of $I_e(t)$ of nonferro-FETs in the early stage. Note that $I_h$ of nonferro-FETs is so low that cannot be measured until breakdown, indicating that there is less hole tunneling back in nonferro-FETs. We call the behavior of FeFETs having $I_e(t)$ increasing with time as a SILC-like behavior, as stress-induced leakage current (SILC) refers to a phenomenon that a leakage current increases with electrical stress. This SILC-like behavior of $I_e(t)$ of FeFETs can be fitted with a power-law function to be $I_e \propto \sqrt{t}$, independent of $V_g$ stress, as displayed in Figures 7C. Increasing gate current over time becomes positive feedback to the damage in the gate insulator, leading to breakdown when $I_e$ is raised to the order of A/cm$^2$. The $I_e$ and $I_h$ levels that trigger breakdown are almost independent of the stress voltage $V_g$.

Time-to-breakdown $t_{BD}$ under CVS are summarized in Figures 7D and 7E for nonferro-FETs and FeFETs, respectively. Not only the breakdown at lower $V_g$ than nonferro-FETs but also $t_{BD}$ more sensitive to $V_g$ can be observed for FeFETs, with $t_{BD}$ of approximately $10^3$ s at $V_g = 3.75$ V reduced to approximately $10^{-1}$ s at $V_g = 4.2$ V. The results of charge-to-breakdown $Q_{BD}$ for electrons $Q_e = \int I_e(t)dt$ and holes $Q_e = \int I_h(t)dt$ are summarized in Figures 7F and 7G for non-ferro FETs and FeFETs, respectively. An obvious difference in the $Q_{BD}$-$V_g$ properties in FeFETs and nonferro-FETs can be observed. While the total electron fluence $Q_e$ of nonferro-FETs at which the breakdown of HfO$_2$ gate insulators occurs has only a weak dependence on stress voltage (note that $Q_h$ could not be extracted as $I_h$ was too low), the total electron $Q_e$ and hole fluences $Q_h$ at which FeFETs reach breakdown vary in a wide range, implying that the total fluence is not a factor that is responsible for the trigger of breakdown of HZO insulators in FeFETs. Figure 7H shows the ratio of $Q_e/Q_h$ at different stress voltages. It is interesting that the electron-to-hole ratio of $Q_{BD}$ of FeFETs is almost constant independent of stress voltage. This behavior is remarkably different from conventional SiO$_2$-gate MOSFETs, where the hole fluence $Q_h$ triggers gate dielectric breakdown and the $Q_e/Q_h$ ratio is not a constant (Chen et al., 1986; Schuegraf et al., 1994a). This finding indicates that the gate dielectric breakdown mechanism in FeFETs should be different from SiO$_2$-gate MOSFETs. We could not compare with nonferro-FETs as $Q_h$ was below the detection limit, so further investigation of the $Q_e/Q_h$ ratio in nonferro-FETs is needed to specify whether or not the constant $Q_e/Q_h$ ratio is a unique feature of FeFETs. Further studies of what physical parameters trigger the breakdown of HZO insulators in FeFETs would provide a clearer understanding of the interaction between the leakage current and gate dielectric breakdown event in FeFETs.

We have observed from Figure 7B that gate leakage increases with stress time, as similar to a SILC-like behavior. Here, we investigate the device behavior during the increase of gate leakage current. Figures 8A and 8C show the $I$-$V_g$ characteristics before and after a CVS at 4 V for 10 s shown in Figure 8B. Although $I_e(t)$ and $I_h(t)$ increase by approximately 100 times during the 10-s CVS test, it is found that an only small change of the $I$-$V_g$ characteristics can be observed after stress. This implies that increases of $I_e(t)$ and $I_h(t)$ in FeFETs are not similar to typical SILC, where increasing current cannot be easily recovered: increasing currents in FeFETs can be recovered after releasing the stress.

This peculiar behavior of the gate leakage current is further investigated by applying interrupt pulses during TDDB tests. Figure 9A displays a voltage waveform when TDDB tests stressed at $V_g$ = 4 V were interrupted by $V_g$ = 0 V for 1 s every stress time of $t_s$. Figures 9B,C show $I_e(t)$ and $I_h(t)$ for each stress cycle when $t_s$ = 10 s (cycles of 4 V for 10 s and 0 V for 1 s). $I_e(t)$ and $I_h(t)$ increase cycle by cycle regardless of interrupts by 0 V, implying that electrical stress keeps accumulated. Figure 9D summarizes the time-to-breakdown $t_{BD}$ (excluding interrupt time at 0 V). $t_{BD}$ independent of interrupt frequency indicates that the interrupts at 0 V have no significant effect on $t_{BD}$. On the other hand, interrupting with negative voltage of $V_g$ = -4 V is different. Figure 9E displays a voltage waveform when interrupted by $V_g$ = -4 V for 1 s every stress time $t_s$. Figures 9F,G illustrate that the SILC-like gate leakage current is recovered after interrupted with $V_g$ = -4 V for 1 s: increasing $I_e(t)$ and $I_h(t)$ are recovered back almost to $I_e(t = 0)$ and $I_h(t = 0)$, respectively, in every cycle. Note that only the current at the first cycle was slightly different because the polarization state of pristine devices is different. This is in agreement with the repeatable $I_g$-$V_g$ and $I_{sub}$-$V_g$ in Figure 5. Due to the recovery of SILC-like behavior, applying negative voltage interruption in this way helps extend the time-to-breakdown $t_{BD}$ by more than an order of magnitude, as summarized in Figure 9H.

### 3.5 Mechanism under voltage stress

The behavior of stress recovery by negative interrupt pulses can be found as well in $HfO_2$ nonferro-FETs, as shown in Figures 10A,B. These facts suggest that although the leakage current and breakdown voltage of $HfO_2$ nonferro-FETs and HZO FeFETs are different in detail due to differences in crystallinity or defect density, the fundamental mechanisms of the breakdown and recovery behavior should be generally similar in $HfO_2$-based materials, for instance, same type of defect generation.

Considering the above findings, we propose the mechanism under high $V_g$ stress, shown in Figure 11. Typically, SILC as well as noisy gate leakage current (PBD; progressive breakdown) under electrical stress before hard breakdown are attributed to the generation of defects such as oxygen vacancies (Olivo et al., 1988; Rofan et al., 1991; DiMaria et al., 1995; Degraeve et al, 1995). On the other hand, the recovery and repeatable behavior of apparently degraded gate leakage currents observed in FeFETs suggests that the defect redistribution should be the main contribution of apparently degraded characteristics rather than the generation of new defects. These defects are redistributed again after applying an opposite voltage pulse, recovered to the condition close to the initial one before stress. This model is supported by the fact that oxygen vacancies can move during the voltage cycling (Pešić et al., 2016; Florent et al., 2018b). However, if the stress is large enough for defects to move to the condition that triggers hard breakdown, suddenly increasing current generates a huge density of defects, which forms a permanent conduction path and results in the failure of the device. Then, the recovery is no longer available for devices that reach the breakdown condition.

Such a memory operation that the polarization states are frequently switched in a bipolar manner can help extend the device lifetime in terms of breakdown failure. In other words, not only the improvement in the material aspect but also choosing an appropriate memory operation is important for the reliability of FeFETs. Whereas bipolar operation is favorable to improving the breakdown-limited endurance, the memory-window-limited endurance has been reported to have the opposite behavior: memory window narrowing is degraded in a bipolar operation faster than in a unipolar operation (Yurchuk et al, 2014). These findings address that the ideal writing operation on the aspects of breakdown and MW narrowing are different. Thus, the endurance tests for evaluating the real lifetime should be carefully designed. Conventional endurance tests of FeFETs using bipolar stress evaluates only one aspect of device endurance, resulting in underestimation of gate dielectric breakdown and overestimation of MW narrowing.

## 4 Conclusion

We investigated the behavior of stress-induced degradation and gate dielectric breakdown in FeFETs with ferroelectric HZO as gate dielectrics on Si substrates. It was observed that gate dielectric breakdown in FeFETs is dominated by the breakdown in the HZO layer, not in the IL. Increasing gate and substrate hole currents under stress, due to the defect movement in HZO, were observed before gate dielectric breakdown occurs. These increasing currents are not a permanent phenomenon: temporary degradation is recovered by applying opposite voltage because of defect redistribution. We found that continuous electrical stress with the same polarity leads to easier hard breakdown, whereas bipolar stress frequently recovers the device distribution and help extend the time-to-breakdown. Because bipolar stress suppresses the breakdown-limited endurance while accelerates the memory window-limited endurance, accurate endurance tests should be carried out to correctly evaluate the endurance characteristics of FeFETs in practical memory operations.

## 5 Conflict of Interest

*The authors declare that the research was conducted in the absence of any commercial or financial relationships that could be construed as a potential conflict of interest.*

## 6 Author Contributions

K.T. and S.T. conceived and proposed the main concepts. K.T. fabricated devices and characterized the electrical properties. K.T., M.T. and S.T analyzed the data and contributed to the in-depth discussion. K.T. and S.T. wrote the manuscript. All authors contributed to the discussions regarding the manuscript.

## 7 Funding

This paper is based on results obtained from a project, JPNP16007, commissioned by New Energy and Industrial Technology Development Organization (NEDO) as well as JST CREST Grant Number JPMJCR20C3 by the Japan Science and Technology Agency (JST).

## 8 Data Availability Statement

The raw data supporting the conclusion of this article will be made available by the authors, without undue reservation.

## 9 References

Böscke, T. S., Müller, J., Bräuhaus, D., Schröder, U., Böttger, U. (2011a). "Ferroelectricity in hafnium oxide thin films," *Appl. Phys. Lett.* 99, 102903. doi: 10.1063/1.3634052

Böscke, T. S., Müller, J. Bräuhaus, D., Schröder, U., Böttger, U. (2011b). "Ferroelectricity in hafnium oxide: CMOS compatible ferroelectric field effect transistors," in Proc. 2011 IEEE International Electron Devices Meeting (IEDM), 547-550, doi: 10.1109/IEDM.2011.6131606

Chen, I. C., Holland, S., Young, K. K., Chang, C., Hu, C. (1986). "Substrate hole current and oxide breakdown," *Appl. Phys. Lett.* 49, 669-671. doi: 10.1063/1.97563

Degraeve, R., Groeseneken, G., Bellens, R., Depas, M., Maes H. E. (1995). "A consistent model for the thickness dependence of intrinsic breakdown in ultra-thin oxides," in Proc. 1995 IEEE International Electron Devices Meeting (IEDM), 863-866, doi: 10.1109/IEDM.1995.499353

DiMaria, D. J., Cartier, E. (1995). "Mechanism for stress-induced leakage currents in thin silicon dioxide films," *J. Appl. Phys.* 78, 3883-3894. doi: 10.1063/1.359905

Dutta, S., Schafer, C., Gomez, J., Ni, K., Joshi, S., Datta, S. (2020). "Supervised learning in all FeFET-based spiking neural network: Opportunities and challenges," *Front. Neurosci.* 14, 634 doi: 10.3389/fnins.2020.00634

Dünkel, S., Trentzsch, M., Richter, R., Moll, P., Fuchs, C., Gehring, O., et al. (2017). "A FeFET based super-low-power ultra-fast embedded NVM technology for 22nm FDSOI and beyond," in Proc. 2017 IEEE International Electron Devices Meeting (IEDM), 485-488, doi: 10.1109/IEDM.2017.8268425

Eitan, B., Kolodny, A. (1983). "Two components of tunneling current in metal-oxide-semiconductor structures," *Appl. Phys. Lett.* 43, 106-108. doi: 10.1063/1.94145

Florent, K., Pesic, M., Subirats, A., Banerjee, K., Lavizzari, S., Arreghini, A. (2018a). "Vertical ferroelectric $HfO_2$ FET based on 3-D NAND architecture: towards dense low-power memory," in Proc. 2018 IEEE International Electron Device Meeting (IEDM), 43-46. doi: 10.1109/IEDM.2018.8614710

Florent, K., Subirats, A., Lavizzari, S., Degraeve, W., Celano, U., Kaczer, B., et al. (2018b). "Investigation of the endurance of FE-$HfO_2$ devices by means of TDDB studies," in Proc. 2018 IEEE International Reliability Physics Symposium (IRPS), 6D. 3.1-6D.3.7. doi: 10.1109/IRPS.2018.8353634

Gong, N., Ma, T.-P. (2018). "A study of endurance issues in $HfO_2$-based ferroelectric field effect transistors: Charge trapping and trap generation," *IEEE Electron Device Lett.* 39, 15-18. doi: 10.1109/LED.2017.2776263

Ichihara, R., Suzuki, K., Kusai, H., Ariyoshi, K., Akari, K., Takano, K., et al. (2020). "Re-examination of $V_{th}$ window and reliability in $HfO_2$ FeFET based on the direct extraction of spontaneous polarization and trap charge during memory operation," in Proc. 2020 Symposia on VLSI Technology and Circuits, TF1.2. doi: 10.1109/VLSITechnology18217.2020.9265055

Jerry, M., Dutta, S., Kazemi, A., Ni, K., Zhang, J., Chen, P.-Y., et al. (2018). "A ferroelectric field effect transistor based synaptic weight cell," *J. Phys. D: Appl. Phys.* 51, 434001. doi: 10.1088/1361-6463/aad6f8

Kim, S. J., Mohan, J., Kim, H. S., Lee, J., Young, C. D., Colombo, L., et al. (2018). "Low-voltage operation and high endurance of 5-nm ferroelectric $Hf_{0.5}Zr_{0.5}O_2$ capacitors," *Appl. Phys. Lett.* 113, 182903. doi: 10.1063/1.5052012

Kobayashi, K., Teramoto, A., Hirayama, M., Fujita, Y. (1995). "Model for the substrate hole current based on thermionic hole emission from the anode during Fowler-Nordheim electron tunneling in n-channel metal-oxide-semiconductor field-effect transistors," *J. Appl. Phys.* 76, 3277-3282, doi: 10.1063/1.358681

Liao, C.-Y., Hsiang, K.-Y., Lou, Z.-F., Tseng, H.-C., Lin, C.-Y., Li, Z.-X., et al. (2022). "Endurance > $10^{11}$ cycling of 3D GAA nanosheet ferroelectric FET with stacked HfZrO$_2$ to homogenize corner field toward mitigate dead zone for high-density eNVM," in Proc. 2022 Symposia on VLSI Technology and Circuits, 393-394, doi: 10.1109/VLSITechnologyandCir46769.2022.9830345

Luo, J., Liu, T., Fu, Z., Wei, X., Yang, M., Chen, L. (2022). "A novel ferroelectric FET-based adaptively-stochastic neuron for stimulated-annealing based optimizer with ultra-low hardware cost," *IEEE Electron Devices Lett.* 43, 308-311. doi: 10.1109/LED.2021.3138765

Matsui, C., Toprasertpong, K., Takagi, S., Takeuchi, K. (2021). "Energy-efficient reliable HZO FeFET computation-in-memory with local multiply & global accumulate array for source-follower & charge-sharing voltage sensing," in Proc. 2021 Symposia on VLSI Technology and Circuits, JFS2-8. doi: 10.23919/VLSICircuits52068.2021.9492448

Migita, S., Ota, H., Yamada, H., Shibuya, K., Sawa, A., Toriumi, A. (2018). "Polarization switching behavior of Hf-Zr-O ferroelectric ultrathin films studied through coercive field characteristics," *Jpn. J. Appl. Phys.* 57, 04FB01, doi: 10.7567/JJAP.57.04FB01

Mulaosmanovic, H., Breyer, E. T., Dünkel, S., Beyer, S., Mikolajick, T., Slesazeck, S. (2021). "Ferroelectric field-effect transistors based on HfO$_2$: a review," *Nanotechnology* 32, 502002. doi: 10.1088/1361-6528/ac189f

Müller, J., Böscke, T. S., Schröder, U., Mueller, S., Bräuhaus, D., Böttger, U., et al., (2012). "Ferroelectricity in simple binary ZrO$_2$ and HfO$_2$," *Nano Lett.* 12, 4318-4323. doi: 10.1021/nl302049k

Müller, J., Polakowski, P., Müller, S., Mulaosmanovic, H., Ocker, J., Mikolajick, T., et al. (2016). "High endurance strategies for hafnium oxide based ferroelectric field effect transistor," in Proc. 16th Non-Volatile Memory Technol. Symp. (NVMTS), 10.1109/NVMTS.2016.7781517

Müller, S., Zhou, H., Benoist, A., Ocker, J., Noack, M., Kuzmanov, G., et al. (2021). "Development Status of Gate-First FeFET Technology," in Proc. 2021 Symposia on VLSI Technology and Circuits, TFS1-5.

Nako, E., Toprasertpong, K., Nakane, R., Takenaka, M., Takagi, S. (2022). "Experimental demonstration of novel scheme of HZO/Si FeFET reservoir computing with parallel data processing for speech recognition," in Proc. 2022 IEEE Symposium on VLSI Technology and Circuits. doi: 10.1109/VLSITechnologyandCir46769.2022.9830412

Ni, K., Sharma, P., Zhang, J., Jerry, M., Smith, J. A., Tapily, K., Clark, R., et al. (2018). "Critical role of interlayer in Hf$_{0.5}$Zr$_{0.5}$O$_2$ ferroelectric FET nonvolatile memory performance," *IEEE Trans. Electron Devices* 65, 2461-2469. doi: 10.1109/TED.2018.2829122

Olivo, P., Nguyen, T. N., Ricco, B. (1988). "High-field-induced degradation in ultra-thin SiO$_2$ films," *IEEE Trans. Electron Devices* 35, 2259-2267. doi: 10.1109/16.8801

Park, M. H., Lee, Y. H., Kim, H. J., Kim, Y. J., Moon, T., Kim, K. D., et al. (2015). "Ferroelectricity and antiferroelectricity of doped thin HfO2-based Films," *Adv. Mater.* 27, 1811-1831. doi: 10.1002/adma.201404531

Peng, H. K., Chan, C. Y., Chen, K. Y., Wu, Y. H. (2021). "Enabling large memory window and high reliability for FeFET memory by integrating AlON interfacial layer," *Appl. Phys. Lett.* 49, 669-671. doi: 10.1063/5.0036824

Pešić, M., Fengler, F. P. G., Larcher, L., Padovani, A., Schenk, T., Grimley, E. D., et al. (2016). "Physical mechanisms behind the field-cycling behavior of HfO$_2$-based ferroelectric capacitors," *Adv. Funct. Mater.* 26, 4601-4612. doi: 10.1002/adfm.201600590

R. Rofan; C. Hu (1991). "Stress-induced oxide leakage," *IEEE Electron Devices Lett.* 12, 632-634. doi: 10.1109/55.119221

Schroeder, U., Park, M. H., Mikolajick, T., Hwang, C. S. (2022). "The fundamentals and applications of ferroelectric HfO$_2$," *Nat. Rev. Mater.* 7, 653-669. doi: 10.1038/s41578-022-00431-2

Schuegraf, K. F., Hu, C. (1994a). "Hole Injection SiO$_2$ Breakdown Model for Very Low Voltage Lifetime Extrapolation," *IEEE Trans. Electron Devices* 41, 761-767. doi: 10.1109/16.285029

Schuegraf, K. F., Hu, C. (1994b). "Metal-oxide-semiconductor field-effect-transistor substrate current during Fowler‑Nordheim tunneling stress and silicon dioxide reliability," *J. Appl. Phys.* 76, 3695-3700, doi: 10.1063/1.357438

Shanware, A., Shiely, J. P., and Massoud, H. Z., Vogel, E., Henson, K., Srivastava, A., Osburn, C., Hauser, J. R., Wortman, J. J. (1999). "Extraction of the Gate Oxide Thickness of N- and P-Channel MOSFETs Below 20A from the Substrate Current Resulting from Valence-Band Electron Tunneling," in Proc. 1999 IEEE International Electron Device Meeting (IEDM), 815-818. doi: 10.1109/IEDM.1999.824274

Sharma, A. A., Doyle, B., Yoo, H. J., Tung, I-C., Kavalieros, J., Metz, M. V., et al. (2020). "High speed memory operation in channel-last, back-gated ferroelectric transistors," in Proc. 2020 International Electron Device Meeting (IEDM), 391-394. doi: 10.1109/IEDM13553.2020.9371940

Takayanagi, M., Takagi, S., Toyoshima, Y., (2001). "Gate voltage dependent model for TDDB lifetime prediction under direct tunneling regime," in Proc. 2001 Symposia on VLSI Technology, 99-100. doi: 10.1109/VLSIT.2001.934968

Tan, A. J., Liao, Y. H., Wang, L. C., Shanker, N., Bae, J. H., Hu, C., et al. (2021). "Ferroelectric HfO$_2$ memory transistors with high-κ interfacial layer and write endurance exceeding $10^{10}$ cycles," *IEEE Electron Device Lett.* 42, 994-997. doi: 10.1109/LED.2021.3083219

Toprasertpong, K., Takenaka, M., Takagi, S. (2019). "Direct observation of interface charge behaviors in FeFET by quasi-static split CV and Hall techniques: Revealing FeFET operation," in Proc. 2019 IEEE International Electron Devices Meeting (IEDM), 570-573. doi: 10.1109/IEDM19573.2019.8993664

Toprasertpong, K., Lin, Z. Y., Lee, T. E., Takenaka, M., Takagi, S. (2020a). "Asymmetric polarization response of electrons and holes in Si FeFETs: Demonstration of absolute polarization hysteresis loop and inversion hole density over 2× $10^{13}$ cm$^{-2}$," in Proc. 2020 Symposia on VLSI Technology and Circuits TF1.5. doi: 10.1109/VLSITechnology18217.2020.9265015

Toprasertpong, K., Tahara, K., Fukui, T., Lin, Z., Watanabe, K., Takenaka, M., et al. (2020b). "Improved ferroelectric/semiconductor interface properties in Hf$_{0.5}$Zr$_{0.5}$O$_2$ ferroelectric FETs by low-temperature annealing," *IEEE Electron Device Lett.* 41, 1588-1591. doi: 10.1109/LED.2020.3019265

Breakdown-Limited Endurance in HZO FeFETs: Mechanism and Improvement Under Bipolar Stress11

Toprasertpong, K., Tahara, K., Hikosaka, Y., Nakamura, K., Saito, H., Takenaka, M. et al. (2022a). "Low Operating voltage, improved breakdown tolerance, and high endurance in Hf$_{0.5}$Zr$_{0.5}$O$_2$ ferroelectric capacitors achieved by thickness scaling down to 4 nm for embedded ferroelectric memory," *ACS Appl. Mater. Interfaces*. 14, 51137-51148. doi: 10.1021/acsami.2c15369

Toprasertpong, K., Nako, E., Nakane, R., Takenaka, M., Takagi, S. (2022b). "Reservoir computing on a silicon platform with a ferroelectric field-effect transistor," *Communications Engineering* 1, 21. doi: 10.1038/s44172-022-00021-8

Toprasertpong, K., Takenaka, M., and Takagi, S. (2022c). "On the strong coupling of polarization and charge trapping in HfO$_2$/Si-based ferroelectric field-effect transistors: overview of device operation and reliability," *Appl. Phys. A* 128, 1114. doi:10.1007/s00339-022-06212-6

Trentzsch, M., Flachowsky, S., Richter, R., Paul, J., Reimer, B., Utess, D., et al. (2016). "A 28nm HKMG super low power embedded NVM technology based on ferroelectric FETs," in Proc. 2016 International Electron Device Meeting (IEDM), 294-297. doi: 10.1109/IEDM.2016.7838397

Weinberg, Z. A., Fischetti, M. V. (1985). "Investigation of the SiO$_2$-induced substrate current in silicon field-effect transistors," *J. Appl. Phys.* 57, 443-452. doi: 10.1063/1.334771

Yan, M.-H., Wu, M.-H., Huang, H.-H., Chen, Y.-H., Chu, Y.-H., Wu, T.-L., et al. (2020). "BEOL-compatible multiple metal-ferroelectric-metal (m-MFM) FETs designed for low voltage (2.5 V), high density, and excellent reliability," in Proc. 2020 International Electron Device Meeting (IEDM), 75-78. doi: 10.1109/IEDM13553.2020.9371916

Yurchuk, E., Mueller S., Martin, D., Slesazeck, S., Schroeder, U., Mikolajick, T., et al. (2014). "Origin of the endurance degradation in the novel HfO$_2$-based 1T ferroelectric nonvolatile memories," in Proc. 2014 IEEE International Reliability Physics Symposium (IRPS), 2E.5.1-2E.5.5. doi: 10.1109/IRPS.2014.6860603

Yurchuk, E., Müller, J., Muller, S., Paul, J., Pesic, M., Bentum, R.v., et al. (2016). "Charge-trapping phenomena in HfO$_2$-based FeFET-type nonvolatile memories," *IEEE Trans. Electron Devices* 63, 3501-3507. doi: 10.1109/TED.2016.2588439

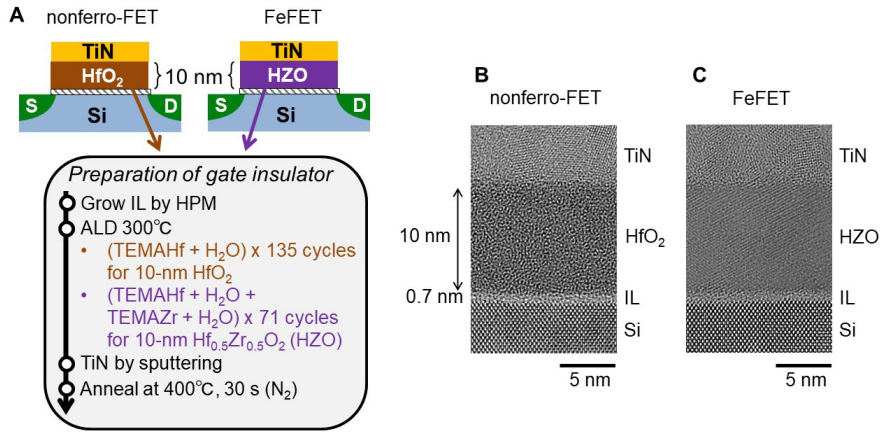

**FIGURE 1 | (A)** Fabrication process flow. TEM images of **(B)** $HfO_2$ nonferro-FET and **(C)** HZO FeFET. HZO was crystallized whereas $HfO_2$ remained amorphous.

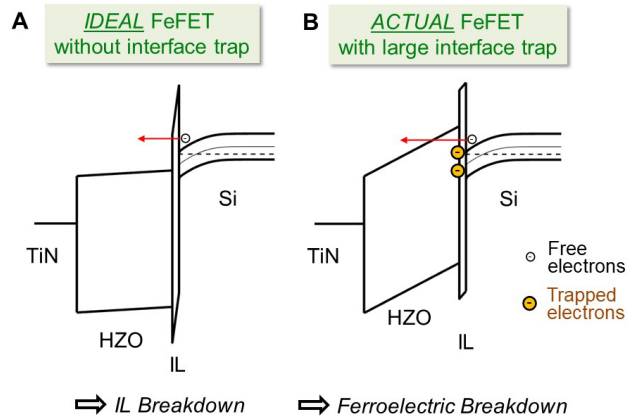

**FIGURE 2 |** Schematic band diagram of HZO/IL/Si gate stack **(A)** when there is no interface charge trapping and **(B)** when there is a large amount of interface charge trapping, where 90% of induced electrons are trapped. Here, the Si band was scaled in the depth direction by 1:100 ratio to make the band bending clear.

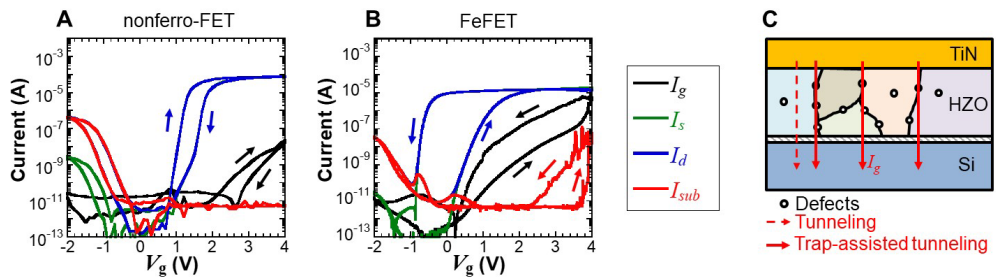

**FIGURE 3 |** Characteristics of $I_g$, $I_d$, $I_s$, and $I_{sub}$ for **(A)** $HfO_2$ nonferro-FET and **(B)** HZO FeFET with $L/W = 10/100$ μm. $I_g$ of a HZO FeFET is around $10^3$ times higher than that of a $HfO_2$ nonferro-FET. Steeply increasing substrate current $I_{sub}$ can be found in FeFETs. **(C)** Leakage current path in ferroelectric HZO gate insulator.

Breakdown-Limited Endurance in HZO FeFETs: Mechanism and Improvement Under Bipolar Stress

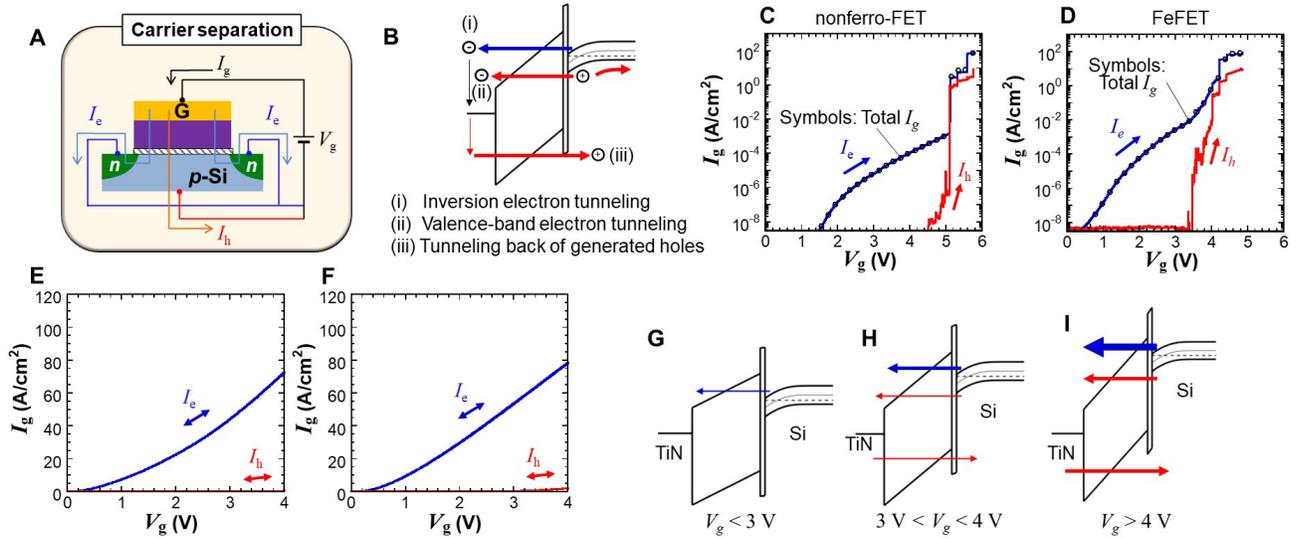

**FIGURE 4 | (A)** Schematic of carrier separation measurement for analyzing gate current. **(B)** Current components in gate current. Inversion electron tunneling flows through S/D, while tunneling of valence-band electrons and generated holes appears as substrate current. Electron-component (blue lines), hole-component (red lines), and total (circle symbols) gate currents of **(C)** nonferro-FET and **(D)** FeFET when $V_g$ was scanned from 0 V until the breakdown point. The electron component dominates the gate current while the hole component rapidly increases near the breakdown voltage. Gate currents after breakdown for **(E)** nonferro-FET and **(F)** FeFET, showing ohmic characteristics. Band diagrams and expected gate current components at **(G)** low $V_g$, **(H)** $V_g < V_{BD}$, and **(I)** $V_g > V_{BD}$.

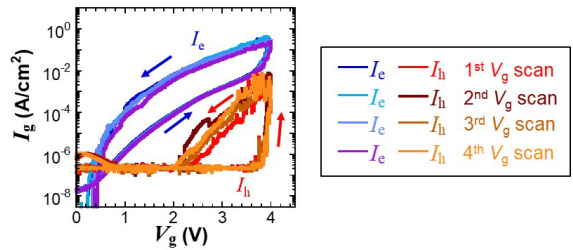

**FIGURE 5 |** Repeatedly measured electron and hole components of $I_g$ in the FeFET. Repeatable current implies that it is not a behavior of permanent trap generation.

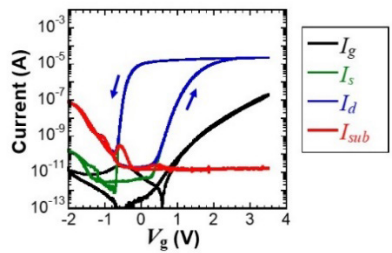

**FIGURE 6 |** Characteristics of $I_g$, $I_d$, $I_s$, and $I_{sub}$ for HZO FeFET with $L/W = 10/100$ μm when the $V_g$ ranged is limited below 3.5 V. No substrate current $I_{sub}$ is observed at positive $V_g$.

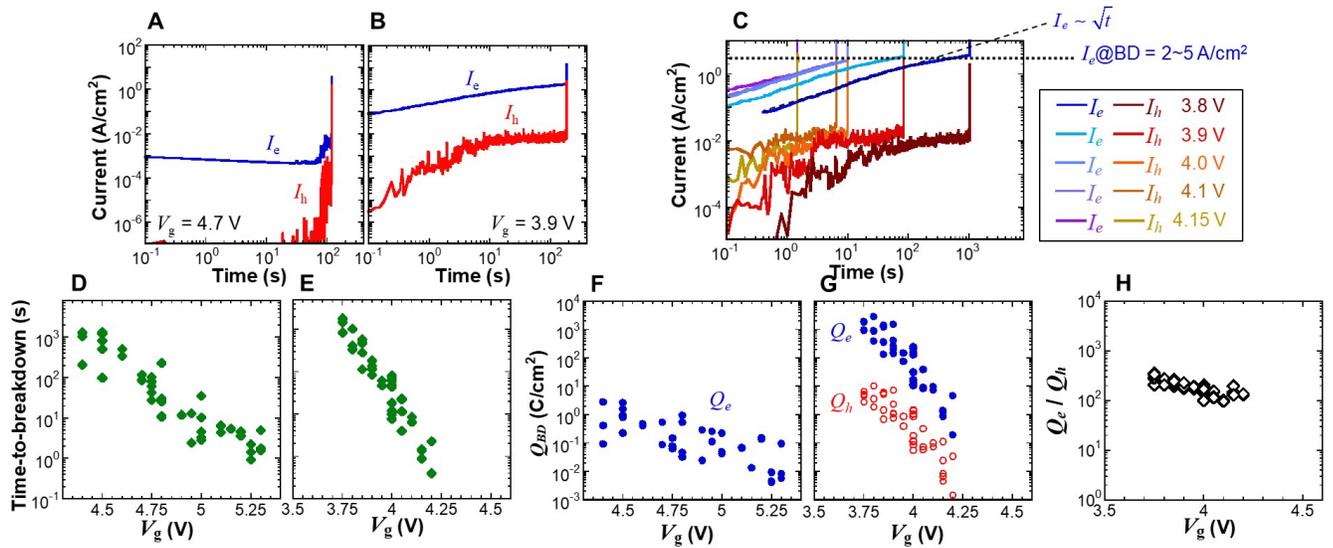

**FIGURE 7 |** TDDB results with carrier separation for **(A)** HfO$_2$ nonferro-FET at $V_g$ = 4.7 V and **(B)** HZO FeFET at $V_g$ = 3.9 V with $L/W$ = 100/100 μm. A SILC-like behavior, with current increasing with stress time, can be observed in FeFETs. **(C)** TDDB of FeFET at different stress voltage $V_g$. Electron current increases with time by approximately $I_e \propto \sqrt{t}$. The electron and hole current levels at breakdown have weak dependency on $V_g$. Time-to-breakdown of **(D)** nonferro-FET and **(E)** FeFET under constant voltage stress. The FeFET has a stronger dependence on $V_g$. Charge-to-breakdown $Q_{BD}$ for **(F)** nonferro-FET and **(G)** FeFET under constant voltage stress. $Q_{BD}$ in the FeFET strongly depends on stressing voltage, whereas $Q_{BD}$ in the nonferro-FET is almost constant. **(H)** $Q_e/Q_h$ ratio at breakdown condition for FeFET.

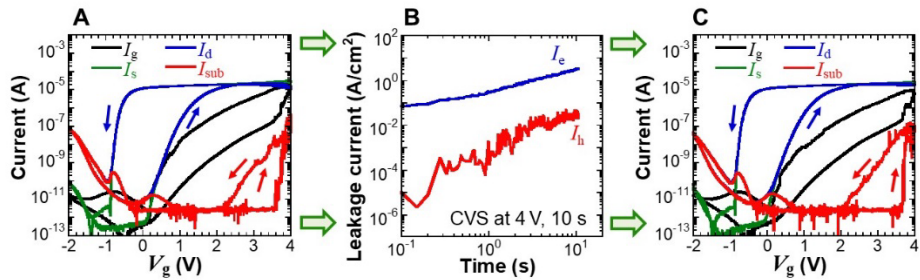

**FIGURE 8 |** **(A)** $I$-$V_g$ characteristics of FeFET before CVS. **(B)** Electron and hole components of gate leakage current under CVS at $V_g$ = 4 V for 10 s. **(C)** $I$-$V_g$ characteristics of FeFET after CVS. Although gate current increases during CVS, it has a negligible effect on $I$-$V_g$ characteristics.

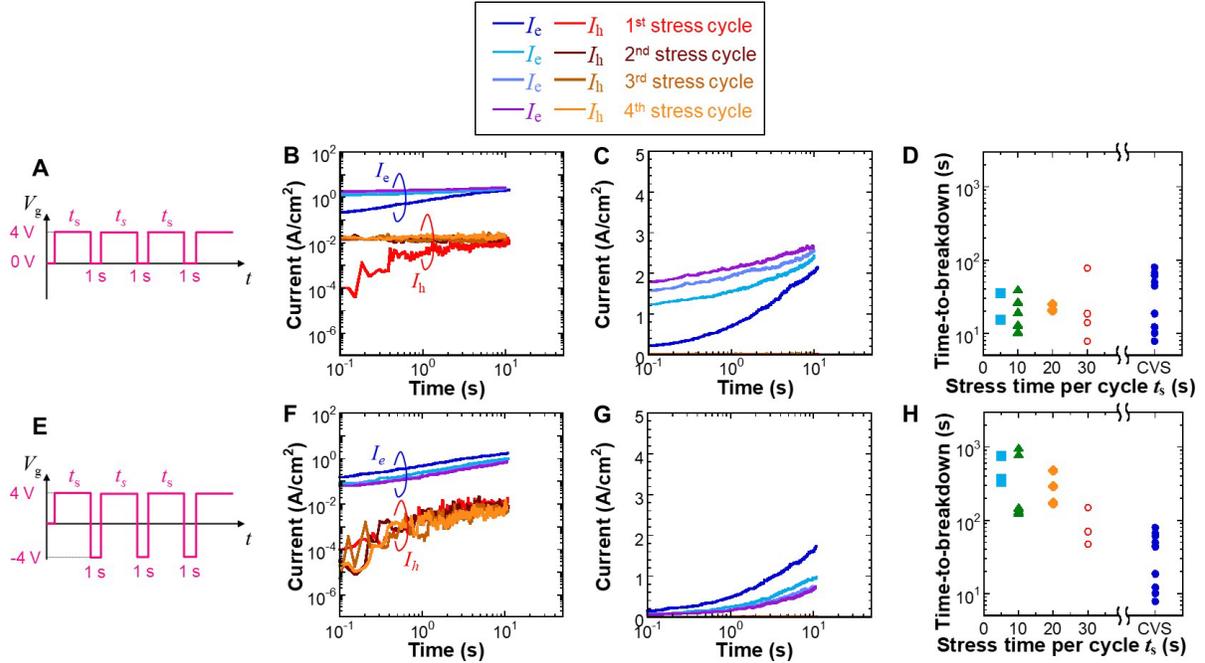

**FIGURE 9 | (A)** Applied voltage scheme with repeating stress of 4 V for time $t_s$ and 0 V for 1 s. **(B,C)** Electron and hole components of gate leakage current at each 4-V stress cycle for $t_s = 10$ s when current is plotted in **(B)** log scale and **(C)** linear scale. Between each stress cycle, tests were interrupted by 0 V for 1 s. **(D)** Total stress time (excluding 0 V interruption duration) before breakdown for different time $t_s$ of 4-V stress. **(E)** Applied voltage scheme when the interrupted voltage is -4 V for 1 s. **(F,G)** Electron and hole components of gate leakage current at each 4-V stress cycle for $t_s = 10$ s, which were interrupted at -4 V for 1 s between cycles, when current is plotted in **(F)** log scale and **(G)** linear scale. **(H)** Total stress time (excluding -4 V interruption duration) before breakdown for different time $t_s$ of 4-V stress.

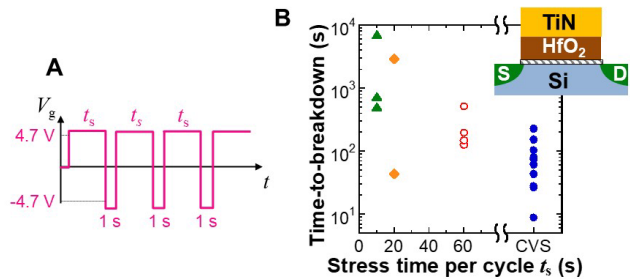

**FIGURE 10 | (A)** Applied voltage scheme with repeating stress of 4.7 V for time $t_s$ and -4.7 V for 1 s. **(B)** Total stress time before breakdown of $HfO_2$ nonferro-FETs. CVS indicates experiments without recovery pulses.

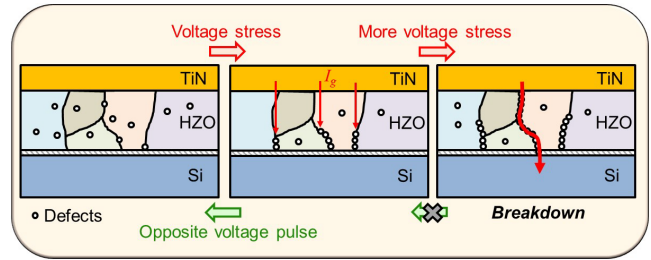

**FIGURE 11** | Mechanism under electrical stress. The SILC-like behavior is attributed to the redistribution of defects rather than permanent defect generation as recovery is observed. Too much stress will trigger breakdown.